\documentclass[10pt]{article}

\usepackage[OE]{express}

\begin{document}
\title{UV-sensitive superconducting nanowire single photon detectors for integration in an ion trap}

\author{D. H. Slichter,\authormark{1,*} V. B. Verma,\authormark{2} D. Leibfried,\authormark{1} R. P. Mirin,\authormark{2} \\ S. W. Nam,\authormark{2} and D. J. Wineland\authormark{1}}

\address{\authormark{1}Time and Frequency Division, National Institute of Standards and Technology, 325 Broadway MC 847, Boulder, CO 80305, USA\\
\authormark{2}Applied Physics Division, National Institute of Standards and Technology, 325 Broadway MC 686, Boulder, CO 80305, USA}

\email{\authormark{*}dhs@nist.gov}


\begin{abstract*}
We demonstrate superconducting nanowire single photon detectors with $76\pm 4$\% system detection efficiency at a wavelength of 315 nm and an operating temperature of 3.2 K, with a background count rate below 1 count per second at saturated detection efficiency.  We propose integrating these detectors into planar surface electrode radio-frequency Paul traps for use in trapped ion quantum information processing.  We operate detectors integrated into test ion trap structures at 3.8 K both with and without typical radio-frequency trapping electric fields.  The trapping fields reduce system detection efficiency by 9\%, but do not increase background count rates.\\
\end{abstract*}



\section{Introduction}

Superconducting nanowire single photon detectors (SNSPDs) are a versatile class of photon-counting detectors exhibiting near-unity detection efficiencies, fast response times, low timing jitter, and very low dark counts over a broad range of wavelengths \cite{Natarajan2012, Marsili2012, Rosenberg2013, Verma2015}.  These properties have made SNSPDs attractive for experiments in quantum optics and quantum information, including recent strong loophole-free tests of local realism, whose success depended crucially on the high detection efficiency and fast response time of the SNSPDs \cite{Shalm2015}.  Current research is also moving beyond single standalone SNSPDs to more complex integrated devices such as larger-scale SNSPD arrays \cite{Allman2015} or SNSPDs integrated with on-chip nanophotonics \cite{Pernice2012, Reithmaier2013, Kahl2015, Najafi2015}.  SNSPDs made from amorphous superconductors such as WSi or MoSi do not require lattice-matched crystalline substrates, easing integration with other technologies and improving device yields \cite{Marsili2012, Verma2015, Verma2014}.  

Most applications of SNSPDs have been in the infrared or at visible wavelengths, although detection at wavelengths down to 370 nm has been demonstrated previously \cite{Verevkin2002, Crain2016, Wollman2016}.  The mechanism of detection, where a photon absorbed by the nanowire creates a local normal-state ``hotspot'' that in the presence of sufficient bias current spreads across the nanowire's width, turning the nanowire briefly resistive, should in principle function for incident radiation of any wavelength capable of creating a suitably large hotspot.  

One application for ultraviolet photon counting is in experiments involving trapped atomic ions, an important technology for quantum information processing.  Quantum state measurement of a trapped ion is typically accomplished by illuminating the ion with laser light resonant with an electric dipole transition and detecting the ion's fluorescence \cite{Leibfried2003}. These measurement transitions are in the ultraviolet or violet for commonly used ion species, including Hg$^+$ (194 nm), Cd$^+$ (227 nm), Mg$^+$ (280 nm), Be$^+$ (313 nm), Yb$^+$ (369 nm), Ca$^+$ (397 nm), and Sr$^+$ (422 nm).  Depending on the illuminated ion's internal state, it will either fluoresce ``bright'', scattering many millions of photons per second in all spatial directions, or not fluoresce, remaining ``dark''.  By collecting and counting some fraction of the scattered photons, it is possible to determine the result of the projective measurement of the ion state.  For high measurement fidelity, the probability distributions of detected counts for the ``bright'' and ``dark'' states (which ideally are Poissonian) must have minimal overlap.  This is achieved by increasing the ``bright'' count rate through a combination of large solid angle for photon collection and high detector quantum efficiency for counting the collected photons, while minimizing the ``dark'' count rate due to intrinsic detector dark counts and stray light.  

Because of their high system detection efficiency (SDE) and very low instrinsic dark count rates, SNSPDs are an ideal candidate for detecting fluorescence photons from trapped ions to achieve high-fidelity quantum state measurement.  In this paper, we report measurements of the sensitivity of MoSi SNSPDs to ultraviolet light at 315 nm (close to the wavelength of $^9$Be$^+$ fluorescence at 313 nm), finding an SDE of $76\pm4$\% when operating at 3.2 K. The background count rate (BCR), which includes both intrinsic detector dark counts and counts due to background photons, was below 1 count per second for bias currents at or below the minimum necessary to saturate the detector efficiency.  For comparison, this represents a factor of two increase in SDE and a reduction by two orders of magnitude in BCR relative to the best commercially available single photon detectors at this wavelength of which we are aware.  

We propose integrating such SNSPDs directly into surface-electrode rf ion traps, which are fabricated on chips using standard lithographic techniques, as a scalable architecture for trapped ion qubit measurement.  As a proof of principle, we fabricate MoSi SNSPDs integrated into test ion trap structures and measure their performance in the presence of realistic rf trapping fields at 3.8 K.  The presence of rf degrades the SDE, but not the BCR, of the trap-integrated detector; a phase-coherent rf tone sent down the SNSPD output line causes the maximum SDE to recover to 91\% of its value with no trapping rf.  We can also operate the same MoSi SNSPD without rf at 4.3 K with a modest penalty in SDE and BCR relative to operation at 3.8 K.

\section{Integrating an SNSPD with a surface electrode ion trap}

Considerable research effort is being directed towards scaling up trapped ion quantum information processors to larger systems with many ions \cite{Monroe2013}.  Surface-electrode ion traps \cite{Seidelin2006a} are a promising technology for this task because they leverage standard semiconductor processing techniques, enabling the fabrication of highly complex multi-zone traps which can host many ions \cite{Chiaverini2005, Amini2010, Stick2010}.  The operation of these traps requires techniques for reading out the states of individual ions in traps containing many ions.  Traditionally, trapped ion state readout has been accomplished by collecting ion fluorescence with a high-numerical-aperture (NA) objective and imaging it onto a photomultiplier tube (PMT) or electron-multiplied charge-coupled device (EMCCD) camera.  Readout fidelity is maximized by choosing an objective with the highest practical NA, and a PMT or EMCCD camera with the best available quantum efficiency and lowest intrinsic dark count rate.  However, there are a number of technical challenges for scaling this method to large numbers of ions.  

With more than one ion in the objective's field of view, a PMT or other non-pixelated detector cannot distinguish which ion is fluorescing.  In this instance, ion-specific readout is accomplished in a time-multiplexed manner, either by illuminating only one ion at a time with a tightly focused laser beam, or by methods involving global microwave or laser pulses in combination with spatial gradients of magnetic or electric fields, which enable the readout transition of the desired ion to be spectrally distinguished from those of other ions \cite{Leibfried1999, Wang2009, Johanning2009a,Warring2013}.  However, the duration of readout performed in this manner increases linearly with the number of ions in the field of view, which hampers scalability to large numbers of ions.  

With an EMCCD camera or other pixelated detector, the fluorescence from different ions can be imaged by the objective onto different sets of pixels, allowing multiple ions to be read out in parallel.  This method requires an imaging objective with high NA and high transmission (to maximize photon collection), large field of view (to image all the ions), low aberrations both on and off axis (to reduce ion fluorescence crosstalk), and appropriate magnification (to allow the ions to be spatially resolved by the detector pixels).  Simultaneous optimization of these parameters is a challenging optical design and fabrication task, especially if the imaging objective is to operate at ultraviolet wavelengths.  Additionally, the diameter of such an objective is typically much larger than its field of view; for traps much larger than the field of view, the use of multiple objectives is hampered by their physical size.  

Precise information on the arrival times of photons at the microsecond time scale can be used to increase the trapped ion readout fidelity further \cite{Langer2006, Myerson2008}.  PMTs typically offer timing resolution in the nanosecond range for individual photon arrivals, while EMCCD cameras are limited by their readout clock frequency and readout noise to a photon arrival timing resolution typically between 100 $\mu$s and 20 ms, depending on the number of active pixels.  

A number of alternatives to the standard ``large objective plus PMT or EMCCD camera'' readout scheme are being pursued.  Optical fibers integrated into a trap can be used to collect ion fluorescence and send it to a detector without requiring an objective \cite{VanDevender2010}, but the fiber integration process is technically demanding. Trap-integrated large-NA spherical mirrors \cite{Merrill2011a}, diffractive lenses \cite{Streed2011, Clark2014}, or diffractive mirrors \cite{Ghadimi2016} can be used to provide focused or collimated output fluorescence beams from each ion, easing the design requirements for the imaging objective.  However, the performance of these optical elements is sensitive to the exact location of the ion relative to the optics, requiring mechanisms for shimming the rf confinement potential for optimal performance \cite{Clark2014, VanRynbach2016}.  Ions can also be placed in a high-finesse optical cavity resonant with the readout transition, into which fluorescence photons will preferentially scatter due to the Purcell effect \cite{Sterk2012}.  This gives increased collection efficiency without increasing the solid angle of the objective optics, but is very sensitive to the alignment of cavity mirrors to each other and to the ion, and may be difficult to scale to large numbers of cavities.  Surface electrode ion traps have been successfully fabricated directly on flat high-finesse cavity mirrors \cite{VanRynbach2016, Herskind2011}, which could reduce the system complexity for cavity integration.  

All of these methods rely on optics and detectors external to the trap structure, as well as mechanical alignment and/or adjustable rf shim electrodes.  In addition, the presence of large areas of exposed dielectric materials (fibers, lenses, or mirrors) near the ions may give rise to time-dependent stray electric fields, which could affect qubit operation and performance.  Standard transparent conductive coatings, such as indium tin oxide (ITO), may not be usable due to poor transmission at many of the relevant ion readout wavelengths.  

An alternative solution for scaling readout is to integrate photon detectors directly into the trap structure \cite{Leibfried2007, Eltony2013, Lekitsch2015}.  Trap-integrated photon detectors require no optics, and their bias and output signals can be routed off the trap chip electrically alongside the electrical connections for the trap electrodes, making them relatively easy to scale up as the trap size grows.  The detectors can be fabricated with precise registration to the trap electrodes on the same substrate, and their collection efficiency is relatively insensitive to small changes in the position of the ion relative to the detector, eliminating the need for mechanical alignment procedures or adjustable rf shim electrodes.  

SNSPDs appear to be prime candidates for use as on-chip photon detectors in surface electrode traps.  Their high SDE and low intrinsic dark counts improve discrimination between ``bright'' and ``dark'' states, and standard SNSPD active areas of $15\times 15 \;\mu$m to $30\times 30 \;\mu$m subtend sufficient solid angle to enable high-fidelity readout given typical ion heights above the trap surface.  SNSPDs fabricated from amorphous superconductors can be made with high yield on commonly used ion trap substrate materials such as silicon \cite{Allman2015}.  

\begin{figure}[tbp]
\centering\includegraphics[width=12cm]{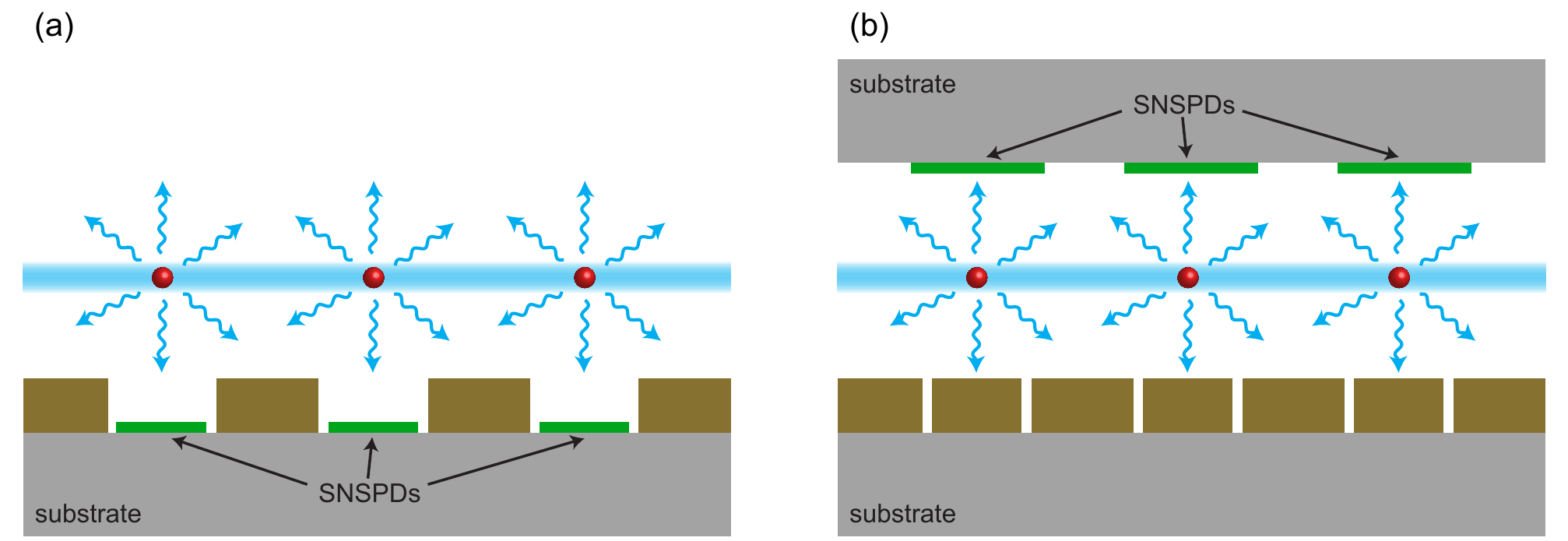}
\caption{\label{detarch}Architecture for scalable readout of trapped ion qubits using integrated SNSPDs.  Surface electrode ion traps (gold electrodes on gray substrate, seen in side view) trap a number of ions (red) above the surface.  These ions are illuminated with a shared readout laser beam (blue) and emit fluorescence photons (blue arrows) depending on their state.  Integrated SNSPDs (green) detect a fraction of the fluorescence photons.  Panel (a) shows a schematic of a trap with integrated detectors, while panel (b) shows a bi-layer ``flip chip'' trap with SNSPDs fabricated on a separate wafer from that which supports the trap electrodes, allowing decoupled optimization of detector geometry and trap electrode geometry.  Wiring to and from the SNSPDs and trap electrodes can be routed on one or more metal layers (not shown for clarity).  Fluorescence crosstalk can be reduced by increasing the lateral distance between ions, reducing the ion height, and/or by using tall electrodes to shield the SNSPDs from neighboring ions, as in (a).  Tall shielding structures could also be implemented in (b), but are not shown.  Typical ion heights above the trap electrodes are 30 to 75 $\mu$m in current designs.}
\end{figure}

Figure \ref{detarch} shows a side view of two possible architectures for integration of SNSPDs with surface electrode traps.  These could apply to detectors integrated in a large-scale trap using the so-called ``quantum CCD'' architecture \cite{Wineland1998, Kielpinski2002}, where ions are shuttled between task-specific trapping locations using time-varying potentials, or to detectors indexed to individual rf microtraps in a two-dimensional array intended for quantum simulations \cite{Schmied2009, Schmied2011}.  Panel (a) shows a standard surface-electrode trap with SNSPDs in multiple ``detection zones'', each with a single ion.  The ions are illuminated with a shared readout beam, and the ion fluorescence is detected in parallel in a spatially-resolved manner by the SNSPDs.  Electrical connections for the SNSPDs and trap electrodes (not shown) are routed in on one or more metal layers using standard fabrication techniques \cite{Stick2010}.  The length of the detector array along the beam will be limited by Gaussian beam diffraction.  As a concrete example, for ion heights of 30 to 75 $\mu$m and a beam waist at the focus of 0.4 times the ion height, the Rayleigh length is between $\sim$1.5-9 mm (assuming 313 nm light).   If the detectors are spaced 3 ion heights apart and are within half a Rayleigh length of the beam focus, this corresponds to an array of roughly 15 to 40 detection zones.  By using multiple parallel readout beams, each with its own array of detectors, hundreds of detection zones could be realized in a given trap.  For scaling to arbitrary array size or shape, on-chip integrated photonics \cite{Mehta2016} could be used to deliver readout laser light to detection zones individually.  Panel (b) shows a variant where the SNSPDs are fabricated on a different substrate from the trap electrodes which is mounted in a flip-chip configuration.  This allows the detector geometry to be optimized separately from the trap electrode geometry, and can also improve the ion trap well depth \cite{Schmied2011, Krauth2015}.  This geometry is seldom used in ion traps because it prevents optical access from the top side, which is typically where the readout objective is placed; however, if the readout system is integrated into the trap structure this optical access may not be required.  

The level of readout crosstalk, when fluorescence photons from an ion in one zone strike a detector in a different zone, is dictated by the height of the ions above the trap surface, the area of the detectors, and the distance between detection zones.  For detectors spaced 3 ion heights apart, the crosstalk is 3-4\%.  Detectors which are recessed below the top of the trap electrodes, as in Fig. \ref{detarch}(a), are shielded to some extent from photons emitted by neighboring ions, further reducing crosstalk.  Crosstalk can be mitigated further with post-processing techniques \cite{Burrell2010}, or by using on-chip integrated photonics \cite{Mehta2016} to time-multiplex the illumination of the detection zones such that when a given zone is illuminated, none of its nearest neighbors are.  The time overhead for this technique depends on the number of nearest neighbors for any given detection zone, but is independent of the total number of detection zones.  

Background counts due to scattered light from the readout laser beam, or from the additional repump laser beams required for some ion species, are also an important concern in this architecture.  Recessing the detectors below the electrode surfaces will shield them from low-angle scatter from these beams, but clean Gaussian beam shapes will be essential for low-background operation.  Background counts due to stray infrared repump photons (such as for Ca$^+$ or Sr$^+$) can be mitigated by  embedding the SNSPDs in an appropriate multilayer optical stack, which should enable a reduction in detector sensitivity at the (IR) repump wavelength of one or two orders of magnitude relative to the (UV) readout wavelength.  Other laser beams may be present during experiments as well, for example to perform qubit state manipulation or sideband cooling, and would need to be time-multiplexed with the readout beam to maintain low background count rates.  Since the SNSPDs can be turned on and off rapidly by adjusting the bias current, and do not exhibit afterpulsing, exposure to scatter from other laser beams either before or after the readout is performed should not compromise the performance of the detectors.

For use as integrated detectors in an ion trap, SNSPDs must not suffer excessive performance degradation in the environmental conditions of an ion trap.  Dissipation from the trapping rf in surface electrode traps can present a substantial thermal load, so ion traps are not typically operated below temperatures of 3-4 K, which is close to $T_c$ for MoSi.  In addition, trap-integrated SNSPDs are exposed to strong rf electric fields used for trapping ions.  The results in section \ref{results} indicate that SNSPD performance under these conditions is still useful for trapped ion readout.  

MoSi SNSPDs have been recently used to read out the state of a single $^{171}\mathrm{Yb}^+$ ion as part of a traditional ion fluorescence detection setup, in place of a PMT \cite{Crain2016}.  Fluorescence photons at a wavelength of 369 nm were collected with a large objective lens and coupled via a multimode fiber to the MoSi SNSPD, operated in a standalone cryostat at 1 K.  Using the ion as a calibrated photon source, the authors extracted a SDE for the detector of $69\pm1$\%.

\section{Design and fabrication}

\begin{figure}[tbp]
\centering\includegraphics[width=13cm]{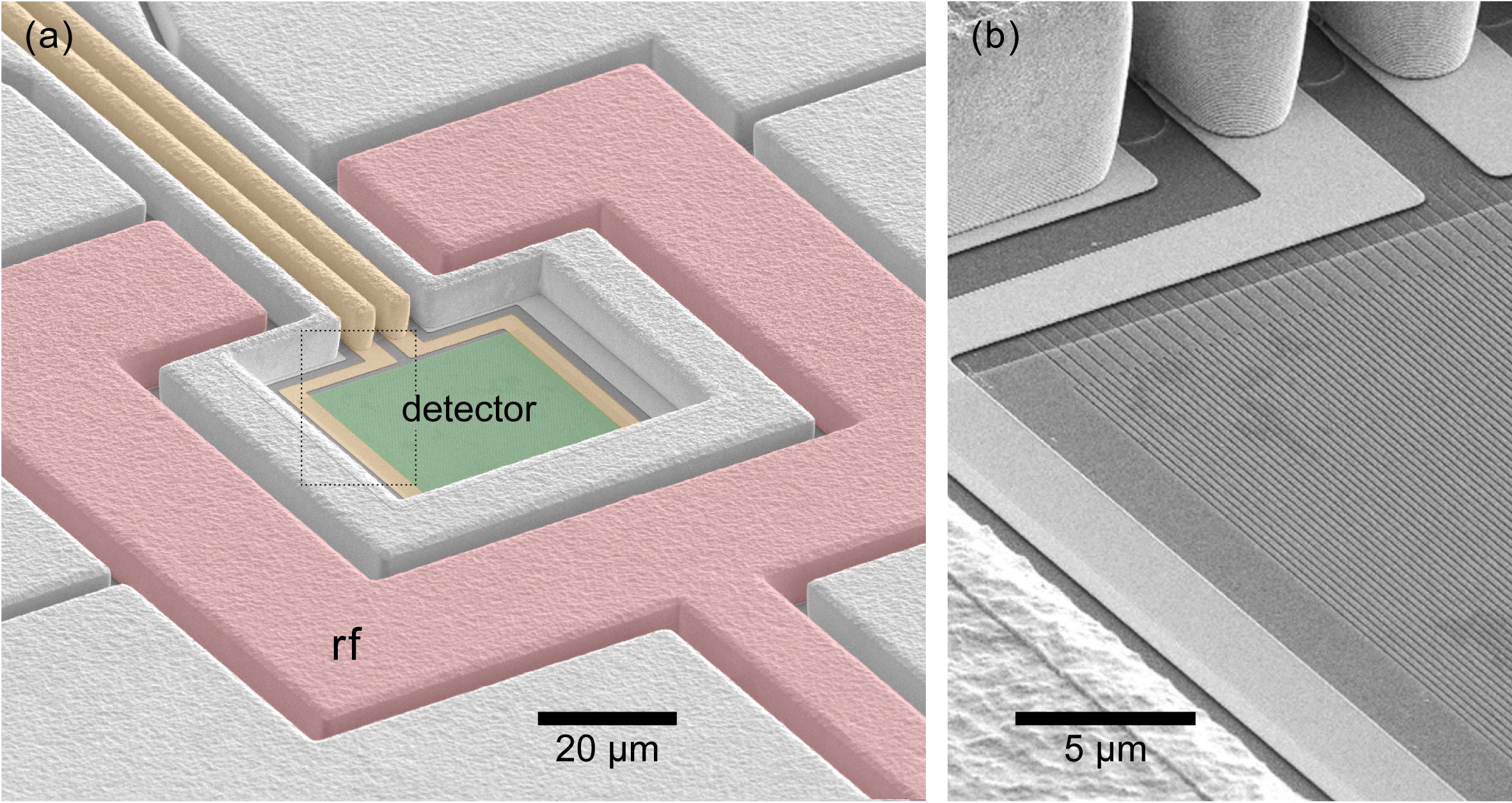}
\caption{\label{trapSEM}Scanning electron micrographs of a trap-integrated SNSPD. Panel (a) is a false-color image, showing the MoSi rectangle (green) containing the nanowire meander, which is connected by electrical leads (gold) to the off-chip bias and readout circuitry. The rf electrode (red) surrounds the SNSPD.  The radio-frequency electric field from this electrode creates a pseudopotential confinement region suitable for trapping a $^9$Be$^+$ ion at a height of 48 $\mu$m above the center of the MoSi rectangle.  The uncolored electrodes are intended as dc shim electrodes; these electrodes were grounded off-chip in our tests.  Panel (b) shows a magnified view of the region inside the dotted rectangle in (a), allowing the nanowire meander to be seen.  The active area of this detector was $30\times30\; \mu$m.}
\end{figure}

Two types of devices were designed and fabricated for the work reported here.  The first design was a stand-alone SNSPD for calibrated tests of detector SDE and dark count performance.  The second design was an SNSPD integrated with additional electrodes to provide electric fields consistent with those required for trapping ions.  All devices were fabricated on 76.2 mm-diameter intrinsic Si wafers (with multiple devices per wafer), on which 65 nm of amorphous SiO$_2$ was deposited by plasma-enhanced chemical vapor deposition (PECVD).  The SiO$_2$ layer and Si substrate form a simple optical stack to enhance photon absorption; the optimal SiO$_2$ thickness assuming normal incidence was calculated using rigorous coupled-wave analysis (RCWA) \cite{Moharam1995}.  Contact pads made of 5 nm Ti followed by 50 nm Au were patterned on top of the SiO$_2$ layer by liftoff. The MoSi film was then deposited by room-temperature dc magnetron sputtering from an alloy target with 75:25 atomic percent MoSi, and optionally capped with 2 nm of amorphous silicon (a-Si) to prevent oxidation of the MoSi film \cite{Verma2015}.  The a-Si cap is thin enough that stray charges trapped on its surface should tunnel to the MoSi layer, thus preventing the a-Si cap from charging up and distorting the potential seen by a trapped ion above it.  The MoSi film was initially etched into rectangles defined by optical lithography, after which the nanowire meander lines were patterned by electron beam lithography and reactive ion etching in an SF$_6$ plasma.  The nanowire width and line pitch of the meander also factor into the photon absorption in the nanowire, and were included in the RCWA simulations.  

After the patterning of the detector nanowires, the silicon substrate of the stand-alone devices was etched through and mounted following the procedure in \cite{Miller2011} to enable mechanical self-alignment of the detector to the core of a butt-coupled optical fiber.  The active area of the stand-alone detectors was $16 \times 16 \;\mu$m, sufficiently larger than the nominal mode field diameter of $\sim2.2\, \mu$m for the single mode fibers and the 3 $\mu$m mechanical alignment tolerance \cite{Miller2011} to ensure full coupling of the light from the fiber to the detector active area.  

The ion trap electrodes in the trap-integrated devices were defined using optical lithography after patterning of the nanowire meanders, and were fabricated by electroplating 7 $\mu$m of Au on top of a Ti/Au seed layer.  After electroplating, the seed layer was removed from the gaps between electrodes by Ar ion milling.  The SNSPDs were protected by photoresist during the electroplating and ion milling steps.  The devices were then diced into square chips with a saw.  The active area of the trap-integrated detectors could be chosen to be any size up to $30 \times 30\; \mu$m by patterning the meander in the appropriate region of the MoSi rectangle.  Figure \ref{trapSEM} shows a scanning electron micrograph of the central portion of a typical trap-integrated device.  The SNSPD meander is patterned in the MoSi rectangle (green), and is contacted at both ends by leads (gold) which connect the meander to the off-chip bias and readout circuitry. The nanowire meander is visible in the detail view in Fig. \ref{trapSEM}(b).  When driven with a radio-frequency potential (and with all other electrodes rf-grounded), the rf electrode (red) creates a three-dimensional pseudopotential well for trapping an ion, located 48 $\mu$m above the center of the MoSi rectangle.  The uncolored dc shim electrodes (grey) can be biased with static potentials to compensate for stray fields at the position of the ion, as well as to tune the directions and frequencies of the secular motional modes of the ion.

\section{\label{results}Experimental results}

We first measured the system detection efficiency (SDE) and background count rate (BCR) of a stand-alone detector with $16 \times 16\;\mu$m active area, butt-coupled to a single-mode step-index UV fiber.   The detector nanowire meander consisted of 90-nm-wide lines on a 190 nm pitch, made from 12-nm-thick MoSi with no a-Si cap layer.  The self-aligned detector/fiber assembly was mounted on the base temperature stage of an adiabatic demagnetization refrigerator (ADR), enabling operation between 250 mK and 3.2 K.  The detector was connected via a coaxial cable to a low-noise output amplifier chain at room temperature with a total gain of 51 dB, followed by a pulse counter or oscilloscope.  The dc bias current $I_b$ was supplied via a room-temperature bias tee between the output amplifiers and the SNSPD.  The illumination for the detector was provided by an ultraviolet light-emitting diode (LED) operating at room temperature, with a center wavelength of 315 nm and full-width at half-maximum (FWHM) of 10 nm, butt-coupled to the other end of the single-mode fiber.  We performed a room-temperature calibration of photon flux at the fiber output versus LED bias using a PMT with a known quantum efficiency at 315 nm.  A separate fiber, twice as long as the detector-coupled fiber, ran from room temperature to the cold stage of the ADR and back to room temperature.  We used this fiber to calibrate the temperature dependence of the fiber attenuation. We define the SDE as the probability of an incident photon from the fiber generating an observed electrical pulse from the detector, and the BCR as the observed count rate with the LED off.  

\begin{figure}[tbp]
\centering\includegraphics[width=10cm]{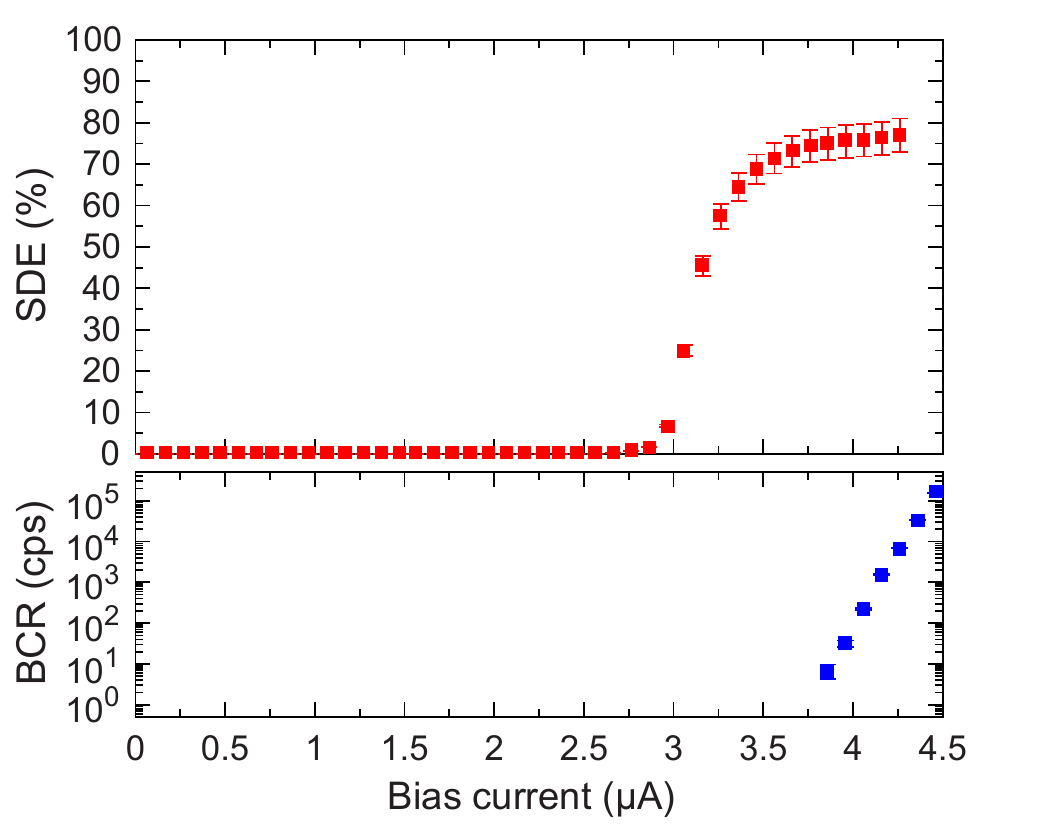}
\caption{\label{standalone}Stand-alone SNSPD performance.  The system detection efficiency (top panel, linear scale) and background count rate in counts per second (bottom panel, logarithmic scale) for a $16\times16\; \mu$m stand-alone detector are plotted versus bias current.  The detector was illuminated with a butt-coupled single-mode fiber and operated at 3.2 K. The switching current $I_{sw}$ was 4.5 $\mu$A for this device at this temperature.  Error bars in the bottom panel are calculated assuming Poissonian statistics.}
\end{figure}

Figure \ref{standalone} shows the performance of a stand-alone detector at 3.2 K.  The detector switching current $I_{sw}$, defined as the maximum current with which the device can be biased without switching to the normal (non-superconducting) state, was 4.5 $\mu$A.  We observed a plateau in SDE for bias currents near $I_{sw}$, with a maximum SDE of $76\pm4$\%.  Modeling using RCWA predicted an SDE of 69\% for this device, in reasonable agreement with the measured value.  The BCR was below 1 count per second (cps) for $I_b \leq 3.8 \,\mu$A.  At $I_b=3.8\, \mu$A, the SDE has already plateaued, measuring $74\pm4$\%.  When operated at 250 mK (not shown in Figure \ref{standalone}), this device exhibited $I_{sw}=14.6 \, \mu$A, with the SDE plateauing at $75\pm4$\% for bias currents above 8.4 $\mu$A.  

For comparison, the PMT used for calibration has an SDE of $38\pm2$\% at 315 nm, the highest at this wavelength of any commercial PMT or EMCCD camera of which we are aware, and a typical BCR of 100 cps.  The performance of the MoSi SNSPD thus represents a factor of two improvement in SDE with two orders of magnitude reduction in BCR relative to the current state of the art.  Further improvements to the SDE of the MoSi SNSPD could be achieved by using a optical stack with a mirror below the SNSPD, as is often done for infrared-optimized SNSPDs \cite{Marsili2012, Verma2015}.  The mirror would need to be made from a material with high reflectivity at 315 nm, such as aluminum, instead of gold or silver; alternatively, a dielectric mirror \cite{Wollman2016, Baek2009} could be used.  

\begin{figure}[tbp]
\centering\includegraphics[width=10.5cm]{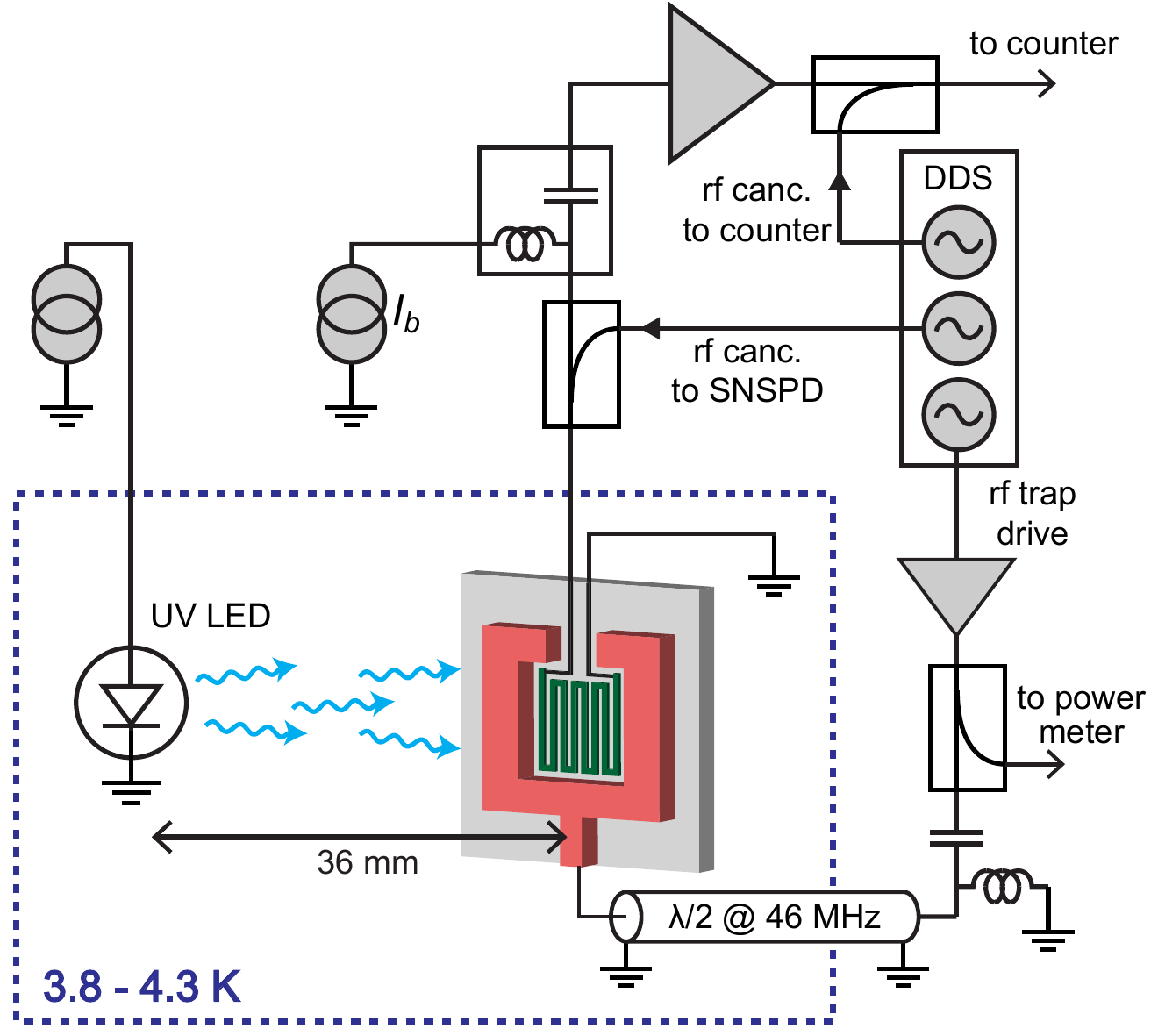}
\caption{\label{rfsetup}Experimental setup for tests with rf drive.  The detector (green meander) is illuminated by a free-space-coupled UV LED.  The detector bias current $I_b$ is applied via a room-temperature bias tee to one detector lead; the other detector lead is grounded.  Output pulses from the SNSPD are amplified and sent to a pulse counter.  The rf electrode (red) is connected to the end of a half-wave coaxial cable resonator, driven by one channel of a multichannel DDS.  Two other channels of the DDS are used to generate phase-coherent rf cancellation tones at the same frequency, which can be sent to the detector and/or to the pulse counter (see text).}
\end{figure}

We also examined the performance of a MoSi SNSPD integrated with an ion trap electrode structure.  The experimental setup is shown in Figure \ref{rfsetup}.  The detector was made from a 7-nm-thick MoSi film with a 2-nm-thick a-Si cap layer, patterned into a meander with 140-nm-wide lines on a 230 nm pitch, with a measured $T_c$ of 5.6 K.  The chip was thermally anchored to the 3 K stage of the ADR, which was augmented with a heater to allow variable temperature operation up to 4.3 K.  The rf electrode was connected to one end of a half-wave coaxial cable resonator with a resonant frequency of 46.23 MHz and a quality factor $Q$ of 27 to provide voltage step-up.  Because of the relatively low resonator $Q$, operation with rf heated up the 3 K stage of the ADR by several hundred millikelvin.  Because $I_{sw}$ and other detector parameters depend strongly on temperature near $T_c$, we adjusted the ADR heater to maintain a constant device temperature of $3.8\pm0.05$ K for all rf drive amplitudes, including rf turned completely off.  The SNSPD was connected to the same bias tee and output amplification chain as for the stand-alone detector, with the addition of two 20 dB directional couplers, one between the amplifier chain and the bias tee and the other following the amplifier chain.  These couplers were used to inject additional rf cancellation tones, one to cancel induced rf currents in the SNSPD and another to cancel rf pickup at the pulse counter.  

All rf tones were generated by a four-channel direct digital synthesizer (DDS), allowing fine tuning of the relative amplitudes and phases of the cancellation signals and ensuring they remained phase coherent with the rf drive.  The rf drive tone was amplified and capacitively coupled to the coaxial resonator, all at room temperature.  A directional coupler immediately before the resonator was used to calibrate the rf drive amplitude.  A room-temperature broadband conical inductor dc-grounded the center conductor of the resonator, and thus the rf electrode on the trap chip.  The rf drive power at the resonator input was approximately +23 dBm.  Most of this power was dissipated in the resonator.

Driving the rf electrode creates a three-dimensional pseudopotential well at a distance of 48 $\mu$m above the center of the MoSi rectangle.   For our applied peak rf amplitude of $V_{pk}=25$ V at $\omega_{rf}/2\pi=46.23$ MHz, and assuming a single $^9$Be$^+$ ion, we calculate the well depth to be 12 meV, with secular motional frequencies of approximately 5 MHz and 9 MHz for the two modes approximately parallel to the plane of the trap, and 14 MHz for the mode approximately normal to the plane of the trap.  These frequencies and well depths are comparable to existing cryogenically operated traps for $^9$Be$^+$, for example in \cite{Brown2011, Wilson2014}. The detector active area subtends 0.9\% of the solid angle seen from the predicted ion position, corresponding to an NA of 0.19.  A detector with the maximum allowable area in the MoSi rectangle ($30\times30 \; \mu$m) would subtend 2.8\% of the solid angle, equivalent to an NA of 0.33.  The maximum possible solid angle available for an integrated detector is dictated by the inside dimensions of the rf electrode (or the distance between the rf electrodes, for a linear trap), which is usually between 1 and 2 times the ion height. Once the shielding electrodes, gaps between electrodes, and the thickness of the trap electrodes are taken into account, an integrated SNSPD can subtend up to 3-8\% of the total solid angle (NA of 0.35-0.55). Suitable large-area SNSPD meanders can be fabricated, at the cost of increased nanowire inductance (which reduces the maximum count rate) and potentially lower fabrication yield.  For comparison, typical ion fluorescence collection objectives for high-fidelity readout subtend 4-10\% of the solid angle (NA of 0.4-0.6) \cite{Noek2013}.

\begin{figure}[tbp]
\centering\includegraphics[width=10cm]{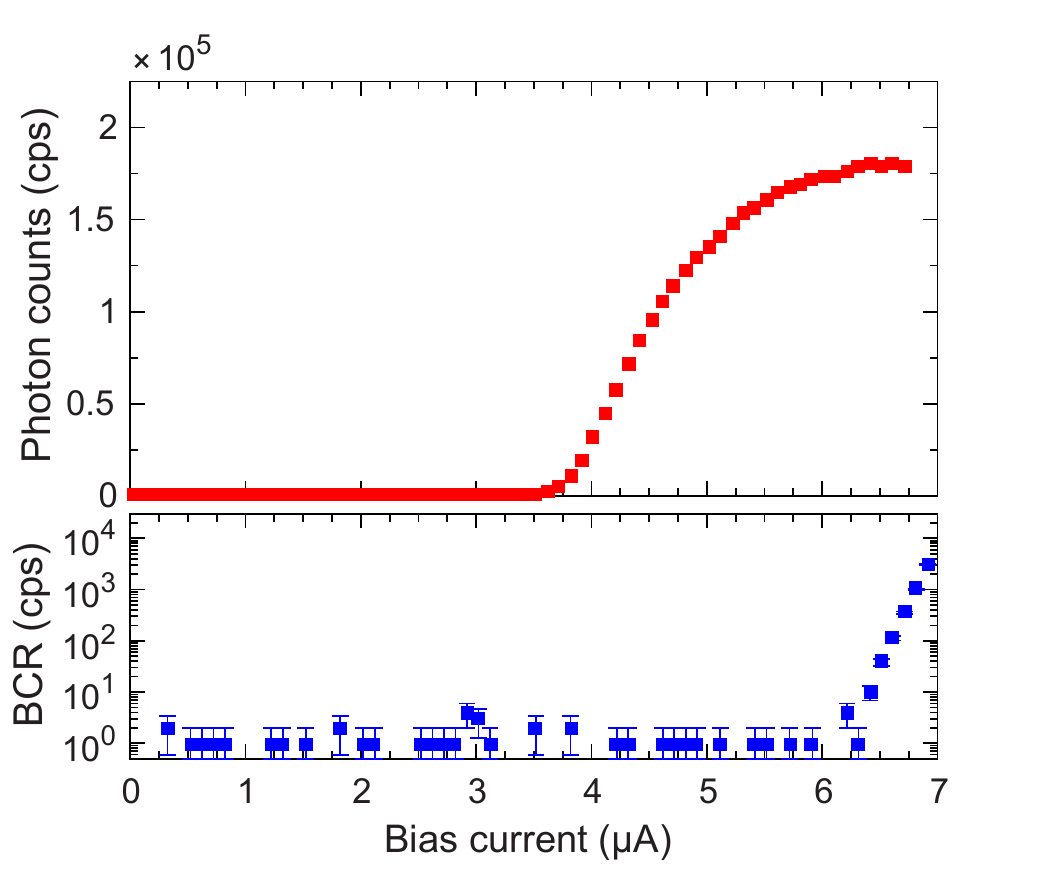}
\caption{\label{lhetemp}Photon count rate (top panel, linear scale) and background count rate (bottom panel, logarithmic scale) for trap-integrated SNSPD operated at 4.3 K with no rf.  The elevated background count rate below $I_b\approx6\;\mu$A may be due to the proximity to $T_c$.   Error bars are calculated assuming Poissonian statistics.  Error bars are smaller than the symbols in the top panel.}
\end{figure}

The illumination for this device was provided by a free-space-coupled UV LED, mounted to the 3 K stage of the ADR approximately 36 mm from the detector.  We could not use a butt-coupled fiber because the presence of the fiber ferrule would distort the rf fields and thus give an inaccurate assessment of the detector tolerance of rf trapping fields.  When operated at 3.8 K, the center wavelength of the LED was 308 nm, with a full width half maximum of 6 nm.  The electrical-to-optical conversion efficiency of the LED was higher at cryogenic temperatures than at room temperature.  We operated the LED with a bias current of 50 $\mu$A and voltage of 5.35 V, which gave count rates from the detectors of approximately $2\times10^5$ counts per second at maximum SDE.  This photon flux was chosen to be similar to the theoretical photon flux on the detector from a single fluorescing ion held by the rf trapping potential.  We present a count rate for this device instead of a calibrated SDE because of the difficulties in accurately calibrating the photon flux at the detector from the free-space-coupled LED; complications arise since photons striking the detector can come either directly from the LED or after one or more reflections from metal surfaces inside the ADR.  If there were no such reflected photons, and only photons traveling directly from the LED struck the detector, we estimate that the photon flux at the detector would be $1.5\pm0.2 \times 10^5$ photons per second.  

\begin{figure}[tbp]
\centering\includegraphics[width=10cm]{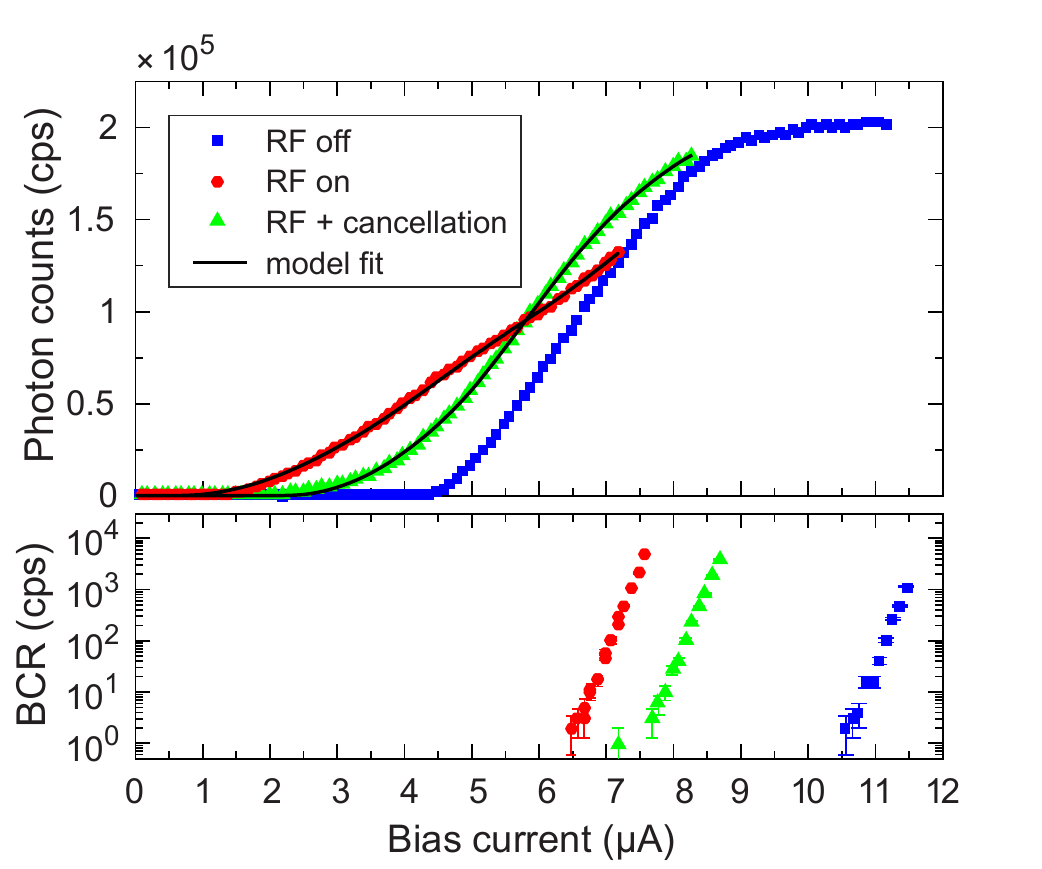}
\caption{\label{rftol}rf tolerance of trap-integrated MoSi SNSPD at 3.8 K.  The photon count rate (linear scale) and background count rate (logarithmic scale) are shown with the rf drive off (blue squares), with the rf drive on with a peak amplitude $V_{pk}=25$ V and frequency $\omega_{rf}/2\pi=46.23$ MHz (red circles), and with both the rf drive as well as a phase-coherent cancellation tone injected down the SNSPD output line via a directional coupler to reduce the amplitude of the rf bias current modulation in the SNSPD (green triangles).  The black lines are fits to the model in Eq. (\ref{simplemodel}).  Error bars are calculated assuming Poissonian statistics.  Error bars are smaller than the symbols in the top panel.}
\end{figure}

We tested the performance of the detector at 4.3 K without rf drive, finding a plateau in the count rate with the LED on, as seen in Figure \ref{lhetemp}.  At this temperature $I_{sw}$ was 6.9 $\mu$A.  The background count rate was roughly 1 count per second, likely due to the proximity to $T_c$.  

We also tested the detector at 3.8 K, both with and without rf trapping fields.  The results are shown in Figure \ref{rftol}.  With the rf drive turned off (blue squares), the measured $I_{sw}$ was 11.5 $\mu$A, with the count rate plateauing for $I_b \gtrsim 9 \; \mu$A.  The photon count rate at the plateau was 13\% higher than for operation at 4.3 K with the same LED bias power.  The BCR was below 1 count per second for $I_b<10.5 \; \mu$A.  The $1/e$ decay time constant of the detector output pulses was $11\pm1$ ns, corresponding to a kinetic inductance of $550\pm50$ nH for the entire meander, or $70\pm7$ pH per square.

Figure \ref{rftol} also shows the detector performance with the rf drive applied (red circles).  The observed switching current was reduced to 7.6 $\mu$A, the photon count rate plateau disappeared, and the maximum photon count rate was only 65\% of the maximum count rate with the rf off.  However, the background count rate remained below 1 cps for $I_b< 6.4 \; \mu$A.  Due to the capacitive coupling between the rf electrode and the SNSPD leads, the output signal from the SNSPD contained detector pulses superimposed on an oscillating rf voltage background of similar amplitude to the detector pulses.  We eliminated this background before sending the signal to the pulse counter by cancelling it with a phase-coherent rf tone injected via the directional coupler after the amplifier chain.  

We conjecture that the observed decrease in switching current and in maximum photon count rate with the rf on is due to rf currents induced in the SNSPD, which modulate the detector bias current at the rf frequency.  These currents arise from the capacitive coupling between the rf electrode and the SNSPD leads and meander.  Since the capacitances between the rf electrode and each SNSPD lead are nominally identical by symmetry, we would expect the rf voltages induced on the two SNSPD leads to be the same, and thus not give rise to a current flow across the SNSPD.  However, this symmetry is broken by the off-chip circuitry, where one lead is shorted to ground, while the other is connected to the 50 ohm input impedance of the amplifier chain.  This asymmetry gives rise to a differential rf voltage between the two SNSPD leads, which is converted by the nonzero rf impedance of the SNSPD to an rf current through the meandered nanowire.  In a separate effect, the oscillating voltage on the rf electrode will alternately draw charges into the meander lines and repel them from the meander lines due to the capacitive coupling between the rf electrode and the SNSPD meander; these moving charges constitute an additional induced rf current to be considered.  The amplitude of this second current will in general be spatially varying across the SNSPD.  

\begin{figure}[tbp]
\centering\includegraphics[width=13cm]{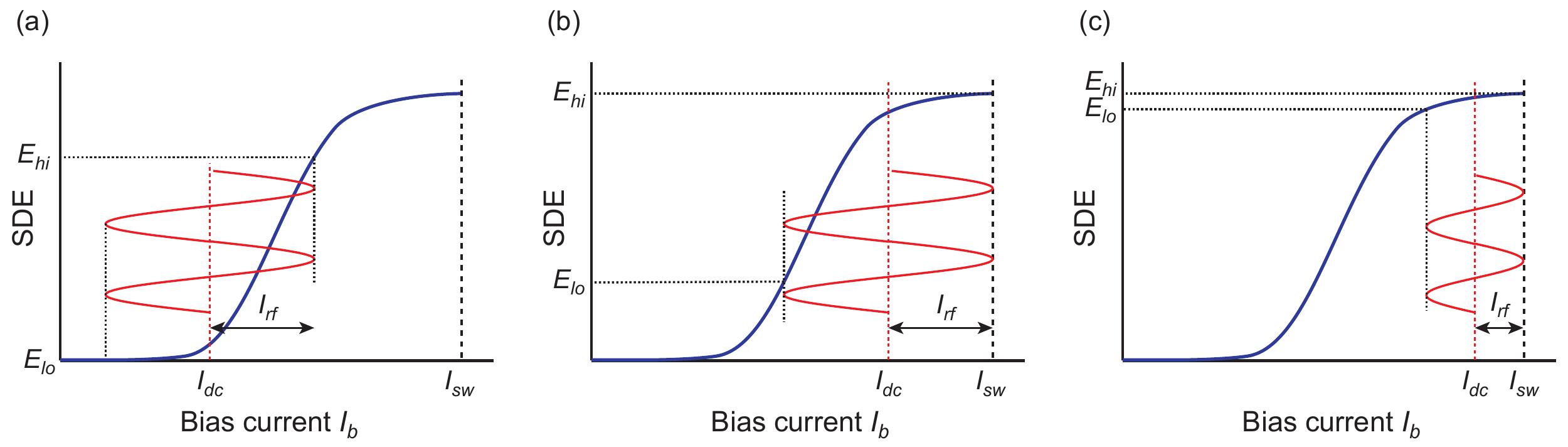}
\caption{\label{rfcartoon}Effect of induced rf currents.  We plot a schematic curve (blue) of a typical SDE versus bias current characteristic for three different combinations of dc bias current $I_{dc}$ and induced rf bias current $I_{rf}$.  Panel (a) shows a large $I_{rf}$ with relatively low $I_{dc}$, panel (b) shows the same $I_{rf}$ with the maximum possible corresponding $I_{dc}$, and panel (c) shows a smaller $I_{rf}$ with the maximum possible corresponding $I_{dc}$.  The red dotted lines indicate the dc bias current, while the solid red sinusoids show the value of the rf-modulated bias current in time.  The effective SDE of the detector, and thus the observed count rate, is given by the time-averaged SDE over one rf cycle.  The maximum and minimum values of the SDE over an rf cycle are denoted $E_{hi}$ and $E_{lo}$, respectively. }
\end{figure}

Figure \ref{rfcartoon} depicts how the rf bias current modulation leads to reduced count rates and smaller observed switching currents.  The three panels show schematic curves of SDE versus bias current (blue lines) for three different combinations of dc bias current $I_{dc}$ (red dotted lines) and induced rf current $I_{rf}$.   The time-dependent bias currents are shown as red sinusoids.  In general, the rf modulation brings the instantaneous bias current to values with both higher and lower SDE than the SDE for dc bias alone.  The effective SDE of the detector with an rf-modulated bias current is given by the time-averaged SDE over one rf cycle.  The instantaneous value of the SDE varies between a minimum of $E_{lo}$ and a maximum of $E_{hi}$ over the course of the rf cycle.

In Fig. \ref{rfcartoon}(a), the rf modulation adds to a relatively small $I_{dc}$, making the SDE considerably higher for some parts of the rf cycle.  The effective SDE with rf modulation will be higher than that from the dc bias alone; compare to the observed count rates in Figure \ref{rftol} for $I_b\approx5 \; \mu$A.  Figure \ref{rfcartoon}(b) demonstrates the mechanism by which the observed switching current is reduced, and why the maximum count rate is lower with the rf drive.  Here $I_{dc}$ has been increased to the maximum possible value given $I_{rf}$; increasing $I_{dc}$ further will cause $I_{sw}$ to be exceeded on every rf cycle. The experimentally measured value of $I_{sw}$ with rf on (7.6 $\mu$A in our experiments) is then this maximum value of $I_{dc}=I_{sw}-I_{rf}$, not the true $I_{sw}$ of the SNSPD (11.5 $\mu$A in our experiments), which is unchanged.  Additionally, the rf modulation causes the instantaneous SDE to be below the plateau value for a substantial portion of each rf cycle, and thus the observed count rate at the maximum $I_{dc}$ will be less than the plateau count rate with no rf drive.  If the amplitude of the induced rf current $I_{rf}$ can be reduced, as shown in Fig. \ref{rfcartoon}(c), the maximum value of $I_{dc}$ will be closer to $I_{sw}$.  If the SDE plateau occurs for a range of bias currents larger than $2I_{rf}$, $I_{dc}$ can be set so that the modulated bias current never leaves the SDE plateau region, and the observed count rate will be close to the plateau count rate with no rf drive.  

Simple numerical simulations of induced rf currents in the SNSPD (see Appendix A) estimated $I_{rf}$ between roughly 3.5 and 6.5 $\mu$A, consistent with the observed reduction in $I_{sw}$ of 3.9 $\mu$A and the mechanism shown in Fig. \ref{rfcartoon}.  To reduce $I_{rf}$ in our sample, and thus increase the SDE at maximum $I_b$, we injected a cancellation tone at $\omega_{rf}$ on the detector output line, towards the detector.  This cancellation tone created an rf current through the SNSPD whose magnitude was set by the SNSPD impedance at $\omega_{rf}$.  By adjusting the phase and amplitude of the cancellation tone relative to the rf drive tone, we created partial destructive interference between this applied current and the currents induced by the rf trap electrode.  The measured photon count rate and background count rate with the cancellation tone are shown in Fig. \ref{rftol} as green triangles.  The observed switching current has increased to 8.7 $\mu$A, and the maximum photon count rate is now 91\% of the maximum count rate with the rf off.  Because we anticipate spatial variation of the induced rf current amplitude and phase in the SNSPD, it should not be possible to completely cancel the induced rf currents at all locations in the nanowire with this technique.  The rf voltage across the SNSPD due to the cancellation tone is roughly $10^5$ times smaller than the rf voltage on the rf electrode, and so we would not expect it to have a significant effect on the motion or confinement depth of a trapped ion.  

We fit the data in Fig. \ref{rftol} to a simple model which assumes that the induced rf currents through the SNSPD have the same phase and amplitude throughout the nanowire.  Expanding the model to allow for a gradient of the induced current amplitude along the meander did not improve the fit further.  The simple model is given mathematically by:

\begin{equation}
\label{simplemodel}
I_b(t)=I_{dc}+I_{rf}\sin(\omega_{rf}t),
\end{equation}
where the two parameters $I_{dc}$ and $I_{rf}$ are allowed to vary in the fitting process.  The fits to our data using Eq. \ref{simplemodel}, both with and without an rf cancellation tone, are shown as solid black lines in Fig. \ref{rftol}.  The fitted value of $I_{rf}$ is 3 $\mu$A without cancellation, in reasonable agreement with numerical simulations.  With cancellation, the fit finds $I_{rf}=1.3 \; \mu$A, consistent with the spatial variation in the induced current along the meander seen in numerical simulations.  We would anticipate that $I_{dc}$ should be given simply by the applied dc bias current, but the fit is substantially improved if $I_{dc}$ is not fixed to this value.  Both fits choose a value for $I_{dc}$ that is approximately 750 nA higher than the applied dc bias current.  This small offset current could potentially arise from rectified rf pickup or interaction of the rf pickup with the bias and amplifier circuitry \cite{Kerman2013}.  

The simulations, fit results, and rf cancellation experiment suggest that the dominant contribution to the induced rf currents in the SNSPD comes from uniform currents generated by capacitive coupling to the SNSPD leads in concert with the asymmetric off-chip termination impedances of those leads.  Designs which reduce the capacitance between the SNSPD leads and the rf electrode, and/or designs which provide bias and readout to the SNSPD while presenting identical impedances to both device leads, should have considerably smaller induced rf currents, with correspondingly improved SDE at maximum dc bias current.  In addition, the simulations indicate that increasing the nanowire inductance (by making the meander longer, with thinner nanowires, for a given active area) also reduces the induced rf currents due to the SNSPD leads.  However, this is counterbalanced by reduced fabrication yields for thinner nanowires, as well as reduced values of $I_{sw}$.  If the induced currents are reduced enough through these passive means, it should be possible to achieve the same SDE both with and without rf (as shown in Fig. \ref{rfcartoon}(c), for example), thus eliminating the need for a cancellation tone injected to the SNSPD.  

\section{Conclusion}

We have demonstrated MoSi SNSPDs with SDE of $76\pm4$\% at a wavelength of 315 nm, with background count rates below 1 cps, operating at a temperature of 3.2 K.  We integrated these SNSPDs into test ion trap structures and demonstrated the feasibility of detecting fluorescence photons from trapped $^9$Be$^+$ ions in the presence of typical rf trapping fields at 3.8 K.  We also demonstrated successful operation of a MoSi SNSPD above liquid helium temperatures.  Our results suggest that integrated SNSPDs are a viable candidate for scaling trapped ion readout in surface electrode traps.  Future work will focus on increasing the detector SDE and reducing induced rf currents in the nanowire, with the aim of detecting fluorescence from a trapped ion with a trap-integrated SNSPD.

\section*{Appendix A: Simulations of induced rf currents}
We performed lumped-element simulations of the SNSPD/rf electrode system with AWR Microwave Office \cite{disclaimer}.  We modeled the SNSPD as a series array of eight identical inductors, with capacitive coupling between each inductor and the rf electrode, and additional capacitive coupling between the SNSPD leads and the rf electrode.  The leads to the off-chip circuitry were modeled as coupled transmission lines.  We used the experimentally determined inductance of the SNSPD to set the total inductance of the array.  We used finite element modeling to estimate the capacitances between the meander and the rf electrode, as well as between the rf electrode and the SNSPD leads.  For our experimental rf drive parameters, the simulations yielded values of the induced current in the meander between 3.5 $\mu$A and 6.5 $\mu$A.  Some of this spread is due to spatial nonuniformity of the induced current amplitude in the meander, as described in the main text; the simulated amplitude of the induced current was about $1.2 \; \mu$A larger at the grounded end of the meander than at the 50-ohm-terminated end of the meander, varying smoothly but not linearly along its length.  The remainder of the spread is due to uncertainty in the rf amplitude calibration and uncertainty in the calculated values of the coupling capacitances.   The phase of the induced current varied smoothly by 40 to 60 degrees over the length of the meander.  The simulated values of induced current amplitude are consistent with the experimentally observed reduction in switching current of 3.9 $\mu$A with the rf turned on, in keeping with the model for switching current reduction as shown in Fig. \ref{rfcartoon}(b) and (c).

\section*{Funding}
Office of the Director of National Intelligence (ODNI) Intelligence Advanced Research Projects Activity (IARPA); NIST Quantum Information Program.

\section*{Acknowledgments}
We thank E. Jordan and J. Bohnet for helpful comments on the manuscript.  All devices were fabricated in the Boulder Microfabrication Facility at NIST.  DHS acknowledges support from an NRC postdoctoral fellowship.  This paper is a contribution of NIST and is not subject to US copyright.

VBV designed and fabricated stand-alone detectors.  DHS and VBV designed and fabricated trap-integrated detectors and performed all experiments.  DHS performed the data analysis and wrote the manuscript.  DL and DJW initiated the proposal for trap-integrated superconducting photon detectors, with input from SWN and RPM.  All authors participated in experimental design and manuscript editing.


\begin{thebibliography}{10}
\bibitem{disclaimer}
Commercial software is identified in this paper for informational purposes only. Such identification does not imply recommendation or endorsement by the National Institute of Standards and Technology, nor is it intended to imply that the software identified is necessarily the best available for the purpose.
\end{thebibliography}
\end{document}